# Ising-Type Magnetic Ordering in Atomically Thin FePS$_3$

*Jae-Ung Lee,[†,∥] Sungmin Lee,[‡,§,∥] Ji Hoon Ryoo,[§,∥] ,Soonmin Kang, [‡,§] Tae Yun Kim,[§] Pilkwang Kim,[§] Cheol-Hwan Park,[§,*] Je-Geun Park, [‡,§,*] and Hyeonsik Cheong[†,*]*

[†]Department of Physics, Sogang University, Seoul 04107, Korea

[‡]Center for Correlated Electron Systems, Institute for Basic Science (IBS), Seoul 08826, Korea

[§]Department of Physics and Astronomy, Seoul National University (SNU), Seoul 08826, Korea

ABSTRACT

Magnetism in two-dimensional materials is not only of fundamental scientific interest but also a promising candidate for numerous applications. However, studies so far, especially the experimental ones, have been mostly limited to the magnetism arising from defects, vacancies, edges or chemical dopants which are all extrinsic effects. Here, we report on the observation of *intrinsic* antiferromagnetic ordering in the two-dimensional limit. By monitoring the Raman peaks that arise from zone folding due to antiferromagnetic ordering at the transition temperature, we demonstrate that FePS$_3$ exhibits an Ising-type antiferromagnetic ordering down to the monolayer limit, in good agreement with the Onsager solution for two-dimensional order-disorder transition. The transition temperature remains almost independent of the thickness from bulk to the monolayer limit with $T_N$ ~118 K, indicating that the weak interlayer interaction has little effect on the antiferromagnetic ordering.

Magnetism has played an important role in advancing our understanding of the quantum nature of materials. Especially, low-dimensional magnetism has been a fertile playground, in which novel physical concepts have been learned and thereby moved the frontiers of the modern understanding of materials science. Most of the three-dimensional magnetic systems, other than some exceptional cases of quantum spin and/or strong frustration, host a magnetic order. On the other hand, fluctuations are so strong and easily destroy stabilization of order parameters in one-dimensional systems as pointed out in the seminal work by Bethe.[1] Two-dimensional (2D) systems, on the other hand, have attracted much attention because the presence or absence of long-range order depends on the type of spin-spin interactions, which themselves compete with intrinsic fluctuations of either quantum and/or thermal nature.

The XXZ Hamiltonian reads[2]

$$H = -J_{XY}\sum_{j\delta}(S_j^x S_{j+\delta}^x + S_j^y S_{j+\delta}^y) - J_I \sum_{j\delta} S_j^z S_{j+\delta}^z, \quad (1)$$

where $J_{XY}$ and $J_I$ are spin-exchange energies on the basal plane and along the c-axis, respectively; $S_j^\alpha$ is the $\alpha$ ($\alpha$ = x, y, or z) component of total spin; and j and $\delta$ run through all lattice sites and nearest neighbors, respectively. The XY model and the Ising model are described by the first term and by the second term on the right hand side of Eq. (1), respectively. According to the Mermin-Wagner theorem,[3] the XY and the isotropic Heisenberg model ($J_{XY} = J_I$) cannot have any long-range order at a finite temperature for 1D or 2D systems. Onsager, on the other hand, demonstrated using an order-converting dual transformation that unlike the 1D system there is a phase transition at a finite temperature in the 2D Ising system.[4] Therefore, ferromagnetic or antiferromagnetic ordering in the 2D limit is possible only in the Ising model. There has been some indirect test of this prediction including the most notable one by Kim and Chan using $CH_4$ molecules adsorbed on graphite.[5] However, despite its fundamental importance, there has been no experimental work using a real 2D magnetic material.

2D van der Waals (vdW) materials could be an ideal system for the study of 2D magnetism.[6] Unfortunately, however, finding suitable vdW materials and producing atomically thin magnetic materials have been a challenge. Although there have been a few reports of producing atomically thin samples of magnetic materials,[7–10] observation of magnetic ordering in the atomically thin limit has been lacking. Transition metal phosphorus trichalcogenides are a new class of magnetic vdW materials. In particular, transition metal phosphorus trisulfides (TMPS$_3$) form in a monoclinic structure with the space group of C2/m. As shown in Figure 1a, the transition metal (TM) atoms (TM=V, Mn, Fe, Co, Ni, or Zn) sit at the 4g position with the local symmetry of 2, forming a honeycomb lattice and are surrounded by six S atoms with trigonal symmetry.[11,12] These S atoms themselves are connected to two P atoms above and below the TM plane like a dumbbell. Because of the S atoms, TMPS$_3$ has a strong van der Waals character and can be easily exfoliated. For example, NiPS$_3$, one of the TMPS$_3$ family, has been successfully exfoliated down to monolayer (1L).[8] Interestingly, the magnetic ground states of these TMPS$_3$'s vary depending on the TM element: FePS$_3$ has an Ising-type transition with $T_N$ = 123 K whereas NiPS$_3$ has a XY-type transition with $T_N$= 155 K and MnPS$_3$ a Heisenberg-type transition with $T_N$= 78 K.[13]

Here, we report on how the Ising-type antiferromagnetic ordering of FePS$_3$ varies as the system approaches the 2D limit. We used Raman spectroscopy to trace the thickness dependence of several strong new peaks due to the antiferromagnetic ordering. We demonstrate that the antiferromagnetic transition of FePS$_3$ survives down to the 1L limit with its transition temperatures almost unchanged by variations in thickness. Raman spectroscopy is widely used for non-destructive characterization of 2D crystals such as graphene and transition metal dichalcogenides.[14,15] In magnetic crystals, low energy excitations such as two magnon scattering appears in the Raman spectrum,[16] or some of the Raman peak positions or intensities change when spins are ordered.[17,18] Since direct measurement of the magnetic properties of atomically thin magnetic materials is difficult, especially in the case of antiferromagnetism, the changes in the Raman spectrum concomitant with a magnetic transition are good alternatives for monitoring magnetic ordering in such materials. In this work, we used polarized Raman spectroscopy to investigate magnetic ordering as a function of temperature in atomically thin $FePS_3$ crystals.

Bulk $FePS_3$ has a monoclinic structure with the factor group $C_{2h}$.[11,12] In contrast, the point group of 1L $FePS_3$ without magnetic order is likely to be $D_{3d}$ which includes three-fold rotational symmetry, assuming that the internal structure of the layer is preserved from bulk to 1L.[19] Figure 1b shows representative Raman spectra of bulk $FePS_3$ at 300 and 80 K. The high frequency modes, $P_3$, $P_4$, $P_5$ and $P_6$, are mostly attributed to the molecular-like vibrations from $(P_2S_6)^{4-}$ bipyramid structures, whereas the low-frequency peaks, $P_1$ and $P_2$, are from vibrations including Fe atoms.[19] Among the two modes comprising $P_4$, $P_{4a}$ phonon mode does not involve Fe movement, whereas Fe vibrates noticeably in $P_{4b}$. The calculated results of corresponding vibrational modes for single-layer $FePS_3$ can be found in Supporting Information (Table S1 and Figures S1–6). Dramatic changes are observed when the temperature is lowered, especially for $P_1$ and $P_2$. These changes are due to spin ordering in Fe atoms as will be explained later. Unlike other 2D anisotropic crystals, each constituent $FePS_3$ layer (without magnetism) has three-fold rotational symmetry, but the stacking of $FePS_3$ in few-layer or bulk $FePS_3$ breaks this symmetry. As shown in Figure 1a, successive layers are stacked slightly shifted in the $a$ axis. The resultant anisotropy is revealed in the polarization-dependence of the Raman spectra measured in parallel polarization ($α=β$). The crystal axes are determined by the single crystal x-ray diffraction (XRD) technique. At 300K,

Figure 1c shows that the position of P$_5$ varies slightly with the incident polarization, which indicates that this peak is a superposition of two peaks with orthogonal polarizations dependences (See Supporting Information Figure S7 for full spectra). When degenerate in-plane modes of an isotropic crystal are subject to uniaxial perturbation, the degeneracy is lifted, and the split peaks have orthogonal polarization dependences.[20] This is a sign of in-plane anisotropy driven by shifted stacking. At 80 K, the pronounced polarization dependence is attributed to magnetic-order-induced anisotropy and also to a possible structural distortion accompanying the magnetic phase transition.[21] In particular, the intensity of P$_1$ shows a clear polarization dependence, with the maximum intensity for the polarization direction of 45º or 135º with respect to the $a$ axis. P$_2$ also shows a similar polarization dependence. These modes are $B_g$-like, whereas P$_{1c}$ and P$_{2a}$, which are $A_g$-like modes, have different polarization.

In order to establish the correlation between Raman spectral changes and the antiferromagnetic phase transition in FePS$_3$, we measured the Raman spectrum of a bulk crystal as a function of temperature as the sample is cooled down through the antiferromagnetic phase transition temperature. Figure 2a shows dramatic changes in the Raman spectrum between 120 and 110 K. This range coincides with the Néel temperature of bulk FePS$_3$. As the crystal undergoes the antiferromagnetic transition, 4 peaks (P$_{1a}$, P$_{1b}$, P$_{1c}$, and P$_{1d}$) emerge from P$_1$ and become sharper as the temperature decreases (see Supporting Information Figure S8). In addition, the intensity of the lowest-frequency peak (P$_{1a}$) shows a dramatic increase below the Néel temperature. We note that P$_1$ at higher temperatures shows an asymmetric and broad line shape, which could be decomposed into two or more Lorentzians. However, because the linewidth of each component at 300 K is larger than the inter-peak separation between those four peaks (at lower temperatures) and we do not have information on the shift of each component with temperature (or the real part of the phonon self energy arising from phonon-phonon and electron-phonon interactions), we were not able to identify the components of $P_1$ at higher temperatures. The intensity of $P_2$ also shows a sudden increase below the Néel temperature.

These dramatic changes can be understood in terms of zone-folding due to magnetic ordering.[22–24] Bulk FePS$_3$ is known to undergo the so-called zigzag type antiferromagnetic ordering[13,25,26] as shown in Figure 2b. The magnetic susceptibility shows a clear sign of an antiferromagnetic phase transition. Note that the change of susceptibility is much larger along the *c* axis than along the *a* or *b* axis, which implies that the spins are aligned along the out-of-plane direction, i.e., the magnetic ordering is of Ising type. In the zigzag-type ordering, the in-plane unit cell doubles in size, and the in-plane first Brillouin zone is halved. We note that the magnetic unit cell is also enlarged along the out-of-plane direction, but in this work, we use the calculated lattice vibrations of a single layer of FePS$_3$ for simplicity. As a result, the *M* point of the Brillouin zone without ordering folds onto the Γ point due to magnetic ordering, and the zone-boundary phonons thus become Raman active (see Figure S9 for illustration of zone folding). In order to understand the experimental results, we calculated the vibrational modes of the ordered phase using density functional theory (DFT) with the frozen phonon method (see Supporting Information Table S1). Although there are quantitative differences in the calculated frequencies, the overall trend is consistent with the experimental results. The lowest-frequency Raman active mode, corresponding to $P_{1a}$, originates from the *M* point of the Brillouin zone without magnetic ordering. This mode would be strongly enhanced only with magnetic ordering due to the zone-folding effect.[19,22–24] In the ferromagnetic or the Néel type antiferromagnetic phase,[27] no such effects are expected because the unit cell size does not change due to magnetic ordering. The next Raman active mode in frequency corresponds to $P_{1b}$ at ~95 cm$^{-1}$ and originates from the Γ point without magnetic ordering. This mode would be observed both with and without magnetic ordering. There are a few other modes with slightly higher frequencies, corresponding to $P_{1c}$ and $P_{1d}$. These modes have mixed nature with components originating from the *M* and Γ points of the unfolded Brillouin zone. Because these modes have partially Γ-like nature, these peaks would appear even without magnetic ordering. Based on these calculations, we interpret the change in the line shape of $P_1$ in the following way. $P_{1a}$ appears only with magnetic ordering. The other components of $P_1$ would have some intensity even above the Néel temperature. We suggest that local fluctuations or disorders would cause broadening of these modes, resulting in the broad line shape of $P_1$ above the Néel temperature. Once there is magnetic ordering, the long-range order would overcome the broadening effect of local fluctuations, and sharp lines would appear. This interpretation is reasonable, given the fact that $P_1$ above the Néel temperature has negligible intensity at the position of $P_{1a}$ after magnetic ordering. The same argument can be applied to $P_2$, which is very weak above the Néel temperature and becomes strong below it. Therefore, we can establish $P_{1a}$ and $P_2$ as the indicators for an antiferromagnetic phase transition. Figure 2c shows the intensities of several Raman peaks as a function of temperature, and the match with the magnetic susceptibility measurement is excellent.

Having established the Raman signatures of antiferromagnetic ordering, we present the data for single-layer $FePS_3$. Figure 3a is an optical contrast image of an exfoliated 1L sample. Figure 3b shows the atomic force microscopy image of the sample showing a thickness of 0.7 nm, which is similar to the interlayer spacing of $FePS_3$, 0.64 nm.[12] Although bulk or few-layer $FePS_3$ has in-plane anisotropy due to shifted stacking, 1L $FePS_3$ without this stacking effect is likely to have three-fold rotational symmetry without magnetic ordering. Figure 3d compares Raman peaks $P_4$ and $P_5$ for several thicknesses at 300K, at incident polarization directions of 45º and 90º. $P_4$, which does not exhibit a polarization dependence in bulk FePS$_3$, shows no anisotropy for all thicknesses. P$_5$, on the other hand, shows a small relative shift between the two polarization directions for all thicknesses except 1L. This confirms that indeed this sample is 1L FePS$_3$ with three-fold rotational symmetry. We note that P$_4$ redshifts as the thickness decreases whereas P$_6$ blueshifts, and so the spacing between these two peaks can be used to estimate the thickness of FePS$_3$. The positions of P$_4$ and P$_6$ as a function of thickness are summarized in Figure S10.

To investigate the thickness dependence of magnetic ordering, we performed polarized Raman measurements on exfoliated samples as a function of temperature. Figure 3c shows the Raman spectra for the sample shown in Figure 3a (See Supporting Information Figure S11–14 for data for other thicknesses). The polarization direction is chosen such that the P$_1$ peaks appear strong. The overall behavior is very similar to the bulk case. Note that the temperature steps between 130 and 100 K are much smaller than in Figure 2a. We see that the intensities of P$_{1a}$ and P$_2$ increase dramatically at low temperature, which is clear evidence for the antiferromagnetic phase transition in 1L FePS$_3$. Figure 3e is a compilation of the temperature dependences of P$_{1a}$ peak intensities for several thicknesses. Remarkably, the Néel temperature has very small, if any, thickness dependence within experimental uncertainties (see Supporting Information Figure S15), which implies that the role of interlayer interaction is negligible for magnetic ordering in FePS$_3$. We note that the linewidth of the P$_{1a}$ peak is somewhat larger for 1L and 2L samples. This is reasonable because in very thin samples, the effect of the disorders in the substrate or the influence of the surface adsorbates would be more pronounced than in thicker samples, and so the extent of the ordering would be shorter than the bulk case. The fact that the Néel temperature has little dependence on the thickness strongly suggests that FePS$_3$ is an Ising-type spin system down to the 2D limit of 1L. To be precise, monolayer FePS$_3$ is a two-dimensional antiferromagnetic anisotropic

Heisenberg system close to the corresponding Ising system. It is known that the former behaves qualitatively the same as the latter.[28]

In conclusion, we investigated the antiferromagnetic phase transition in FePS$_3$ as a function of thickness by monitoring the Raman peaks that result from zone folding due to antiferromagnetic ordering. FePS$_3$ exhibits an Ising-type antiferromagnetic ordering even down to the monolayer limit. The transition temperature is almost independent of the thickness, indicating that the interlayer interaction has little effect on the magnetic ordering. Our study demonstrates that two-dimensional materials can have robust intrinsic magnetism.

**Methods.** *Crystal growth and characterization.* Single-crystal FePS$_3$ was grown by chemical vapor transport using pure Fe, P, and S powders (Alfa Aesar), with purity better than 3N. The starting materials were sealed into a quartz ampoule with pressure under $10^{-2}$ Torr. The temperatures were set at 750 ℃ (hot zone) and 730 ℃ (cold zone) using a horizontal 2-zone furnace and kept at that condition for 9 days. We checked the stoichiometry of all our single crystal samples using a scanning electron microscope (COXI EM30, COXEM) equipped with an energy-dispersive x-ray spectrometer (Bruker Quantax 100). The quality as well as the orientation of the samples were checked by using two x-ray diffractometers including a Laue diffractometer (TRY-IP-YGR, TRY SE) and a single crystal diffractometer (XtaLAB P200, Rigaku), and all our samples were found to form in the C2/m symmetry. The bulk properties of the samples such as the magnetic susceptibility and the heat capacity were measured using a commercial system (PPMS9, Quantum Design, USA) and a SQUID-VSM (MPMS 3, Quantum Design, USA). An antiferromagnetic transition was observed at T$_N$ = 118 K with the Curie-Weiss temperature of $\theta_{CW}$ = –112 K.

*Temperature dependent polarized Raman measurements.* The samples were prepared on Si substrates with a layer of 285-nm $SiO_2$ by mechanical exfoliation from bulk single-crystal $FePS_3$. The atomically thin samples are relatively stable but can be degraded when the samples are exposed in ambient condition for more than a week (See Supporting Information Figure S16). After exfoliation, the samples were kept in a vacuum desiccator to prevent possible degradations. All the measurements were performed in vacuum using an optical cryostat (Oxford Micorostat He2) at temperatures from 77 to 300 K. The 514.5-nm (2.41 eV) line of an Ar ion laser was used as the excitation source. The laser beam was focused onto the sample by a 40× microscope objective lens (0.6 N.A.), and the scattered light was collected and collimated by the same objective. The laser power was kept below 100 μW in order to avoid local heating. The scattered signal was dispersed by a Jobin-Yvon Horiba iHR550 spectrometer (2400 grooves/mm) and detected with a liquid-nitrogen-cooled back-illuminated charge-coupled-device (CCD) detector. Volume holographic filters (Ondax) were used to clean the laser lines and reject the Rayleigh-scattered light. An achromatic half-wave plate was used to rotate the polarization of the linearly polarized laser beam to the desired direction. The analyzer angle was set such that photons with polarization parallel to the incident polarization pass through. Another achromatic half-wave plate was placed in front of the spectrometer to keep the polarization direction of the signal entering the spectrometer constant with respect to the groove direction of the grating.[29] The temperature dependent Raman measurements on bulk crystal (Figure 2) was carried out using a closed-cycle He cryostat. The excitation laser was focused by a spherical lens ($f$=75 mm) to a spot size of ~50 μm. The power of the laser was 1 mW. For the degradation test in ambient conditions (Supporting Information Figure S16), we used a commercial Raman spectrometer system (XperRam Compact, Nanobase) with 532-nm excitation.

*Calculation details.* We obtained the inter-ionic force constants using density functional theory and frozen-phonon method with Quantum Espresso[30] and Phonopy[31] packages. We performed simulations on isolated FePS$_3$ layers by setting the inter-layer distance to 15 Å. We fully relaxed the lattice parameters and the atomic positions. Projector augmented wave method is used to describe ion-electron interactions, with scalar-relativistic effects incorporated. The exchange-correlation energy is described using the Perdew–Burke–Ernzerhof functional.[32] In order to describe the correlation effects of 3d electrons in Fe, we have used 4.2 eV for the Hubbard U parameter of Fe. A $7 \times 7 \times 1$ *k*-point mesh was used, and the kinetic energy cutoff was set to 40 Ry.

FIGURES

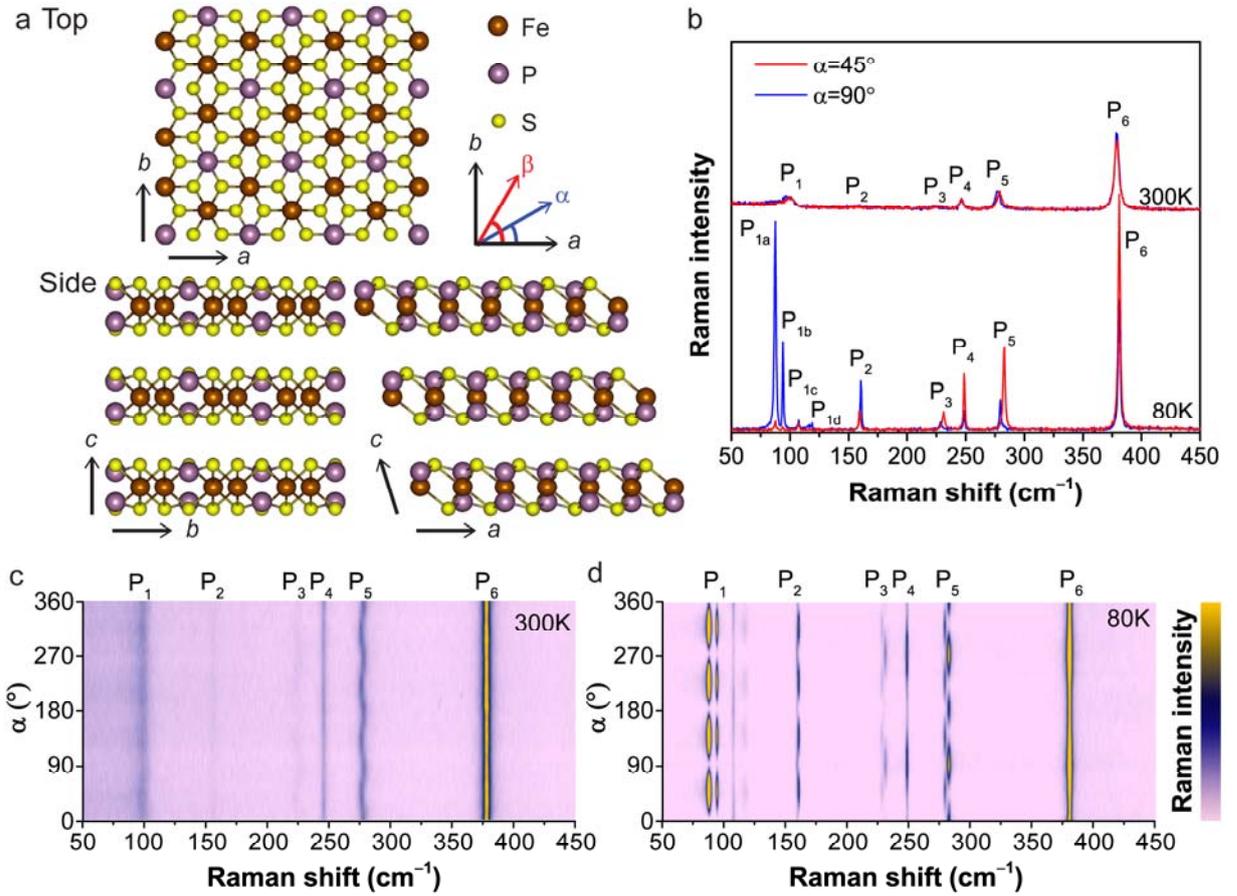

**Figure 1.** (a) Crystal structure of FePS$_3$. Definitions of incident ($\alpha$) and scattered ($\beta$) polarization angles are shown in the top view. (b) Polarized Raman spectra of bulk FePS$_3$ at 80 and 300 K in parallel polarization configuration ($\alpha=\beta$), for $\alpha$ = 45° and 90°, respectively. (c,d) Polarization dependence of Raman spectra at (c) 80 and (d) 300 K.

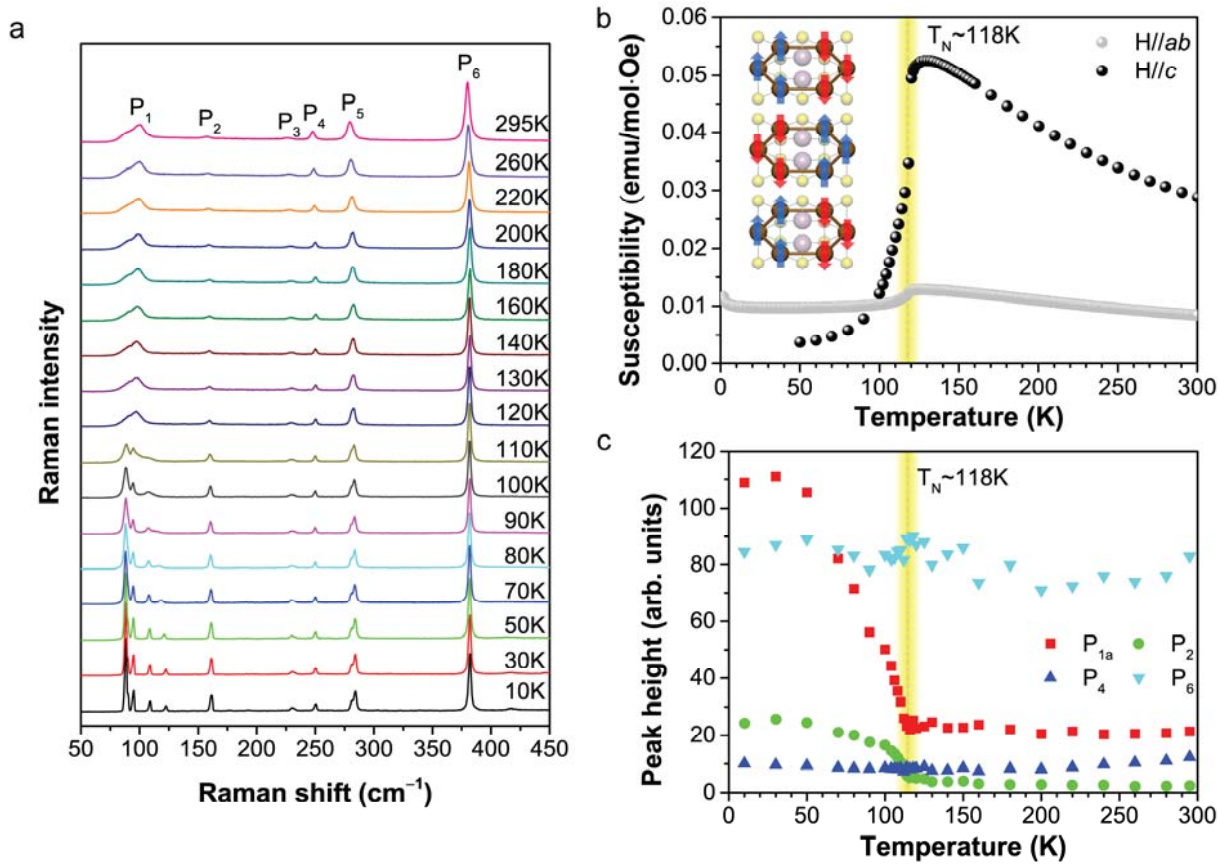

**Figure 2.** (a) Temperature dependence of Raman spectrum for bulk FePS$_3$. (b) Temperature dependence of magnetic susceptibility along *a* or *b* (black spheres) and *c* (grey spheres) axes. (c) Temperature dependence of intensities of several Raman peaks.

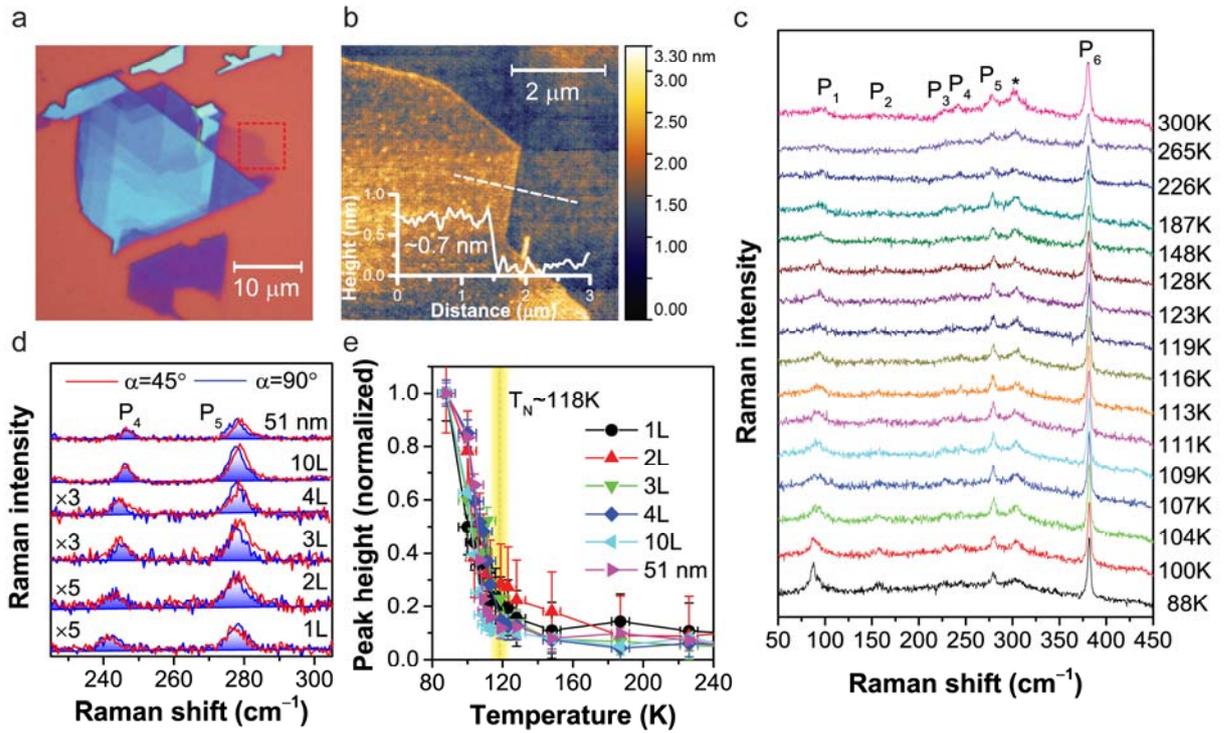

**Figure 3.** (a) Optical contrast and (b) atomic force microscope images of 1L FePS$_3$ on SiO$_2$/Si substrate. (c) Temperature dependence of Raman spectrum of 1L FePS$_3$ with polarization direction of $\alpha=\beta=45°$. * indicates a signal from the Si substrate. (d) Thickness dependence of P$_3$ and P$_4$ with polarization direction of $\alpha=45°$ (red curves) and $\alpha=90°$ (blue curves). (e) Temperature dependence of P$_{1a}$ peak height for different thicknesses (see also Supporting Information Figure S14).

## ASSOCIATED CONTENT

**Supporting Information.** Calculated Raman modes of 1L $FePS_3$ with antiferromagnetic order. Polarized Raman spectra of bulk $FePS_3$ in the extended spectral range. Temperature dependence of $P_1$ components in bulk $FePS_3$. Schematics and examples of zone-folding effect. Peak positions of $P_4$ and $P_6$ and their difference ($P_4-P_6$) as a function of thickness. Representative atomic force microscopy images of exfoliated $FePS_3$. Temperature dependence of Raman spectra for 2L, 3L, 4L, 10L, and 51-nm-thick samples. Temperature dependence of $P_{1a}$ peak height and obtained Néel temperature with different thicknesses. Degradation of 2L $FePS_3$ exposed in the ambient condition. Effect of symmetry on calculated atomic structure and phonons of monolayer $FePS_3$.

## AUTHOR INFORMATION

**Corresponding Authors**

*E-mail: cheolhwan@snu.ac.kr.

*E-mail: jgpark10@snu.ac.kr.

*E-mail: hcheong@sogang.ac.kr

**Author Contributions**

J.-G.P. and H.C. conceived the experiments. S.L. prepared bulk crystals. J.-U.L. performed Raman measurements. S.K. performed x-ray characterizations. J.H.R, T.Y.K., P.K. C.-H.P. carried out calculations. The data were discussed by all the authors, and the manuscript was written through contributions of all authors. All authors have given approval to the final version of the manuscript. ‖These authors contributed equally.

**Notes**

The authors declare no competing financial interest.

ACKNOWLEDGEMENT

This work was supported by the National Research Foundation (NRF) grants funded by the Korean government (MSIP) (Nos. NRF-2016R1A2B300863 and NRF-2013R1A1A1076141) and by a grant (No. 2011-0031630) from the Center for Advanced Soft Electronics under the Global Frontier Research Program of MSIP. Work at IBS-CCES was supported by Institute for Basic Science (IBS-R009-G1). We thank Hyun Cheol Lee for helpful discussions.

Table of Contents Graphic

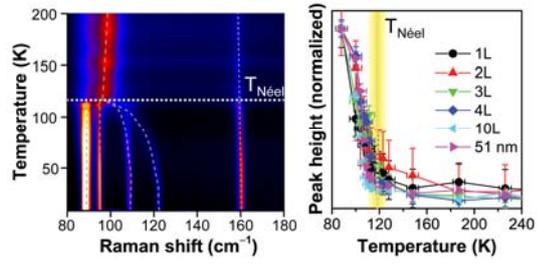

# Supporting Information

# Ising-Type Magnetic Ordering in Atomically Thin FePS$_3$

*Jae-Ung Lee,*[†,∥] *Sungmin Lee,*[‡,§,∥] *Ji Hoon Ryoo,*[§,∥] *,Soonmin Kang,*[‡,§] *Tae Yun Kim,*[§] *Pilkwang Kim,*[§] *Cheol-Hwan Park,*[§,*] *Je-Geun Park,*[‡,§,*] *and Hyeonsik Cheong*[†,*]

[†]Department of Physics, Sogang University, Seoul 04107, Korea

[‡]Center for Correlated Electron Systems, Institute for Basic Science (IBS), Seoul 08826, Korea

[§]Department of Physics and Astronomy, Seoul National University (SNU), Seoul 08826, Korea

**Contents:**

**I. Additional data**

**Table S1.** Calculated Raman modes of 1L FePS$_3$ with antiferromagnetic ordering.

**Figure S1.** Calculated lowest-frequency Raman active modes.

**Figure S2.** Calculated Raman active modes corresponding to $P_2$.

**Figure S3.** Calculated Raman active modes corresponding to $P_3$.

**Figure S4.** Calculated Raman active modes corresponding to $P_4$.

**Figure S5.** Calculated Raman active modes corresponding to $P_5$.

**Figure S6.** Calculated Raman active modes corresponding to $P_6$.

**Figure S7.** Polarized Raman spectra of bulk $FePS_3$ in the extended spectral range.

**Figure S8.** Temperature dependence of $P_1$ components in bulk $FePS_3$. Red dashed curves are guide to eyes.

**Figure S9.** Schematics and examples of zone-folding effect.

**Figure S10.** Peak positions of $P_4$ and $P_6$ and their difference ($P_6-P_4$) as a function of thickness.

**Figure S11.** Representative atomic force microscopy images of exfoliated $FePS_3$.

**Figure S12.** Temperature dependence of Raman spectra for 2L and 3L samples.

**Figure S13.** Temperature dependence of Raman spectra for 4L and 10L samples.

**Figure S14.** Temperature dependence of Raman spectrum for 51-nm-thick sample.

**Figure S15.** Temperature dependence of $P_{1a}$ peak height and obtained Néel temperature with different thicknesses.

**Figure S16.** Degradation of 2L $FePS_3$ exposed in the ambient condition estimated from Raman spectrum.

## II. Effect of symmetry on calculated atomic structure and phonons of FePS$_3$

**Figure S17.** Lowest-energy stable phonon modes of monolayer FePS$_3$; the point group is constrained to $C_{2h}$.

**Figure S18.** Unstable phonon mode of monolayer FePS$_3$ when the point group is constrained to $C_{2h}$.

# I. Additional Data

**Table S1.** Calculated Raman modes of 1L $FePS_3$ with antiferromagnetic ordering.

| | Raman shift ($cm^{-1}$) | | Raman shift ($cm^{-1}$) | | Raman shift ($cm^{-1}$) |
|---|---|---|---|---|---|
| $P_{1a}$ | 67.6  | | 195.3 | $P_{5c}$ | 271.6 |
| $P_{1b}$ | 81.7  | | 197.4 | $P_{5d}$ | 271.7 |
| $P_{1c}$ | 90.4  | $P_{3a}$ | 216.7 | $P_{6a}$ | 364.7 |
| $P_{1d}$ | 94.7  | $P_{3b}$ | 218.0 | $P_{6b}$ | 369.3 |
| $P_{1e}$ | 101.5 | $P_{3c}$ | 219.2 | | 536.6 |
|          | 119.7 | $P_{3d}$ | 222.8 | | 539.9 |
| $P_{2a}$ | 145.0 | $P_{4a}$ | 233.1 | | 540.0 |
| $P_{2b}$ | 148.5 | $P_{4b}$ | 234.8 | | 543.8 |
| $P_{2c}$ | 152.9 | $P_{5a}$ | 260.2 | | 544.0 |
| $P_{2d}$ | 161.1 | $P_{5b}$ | 265.1 | | |

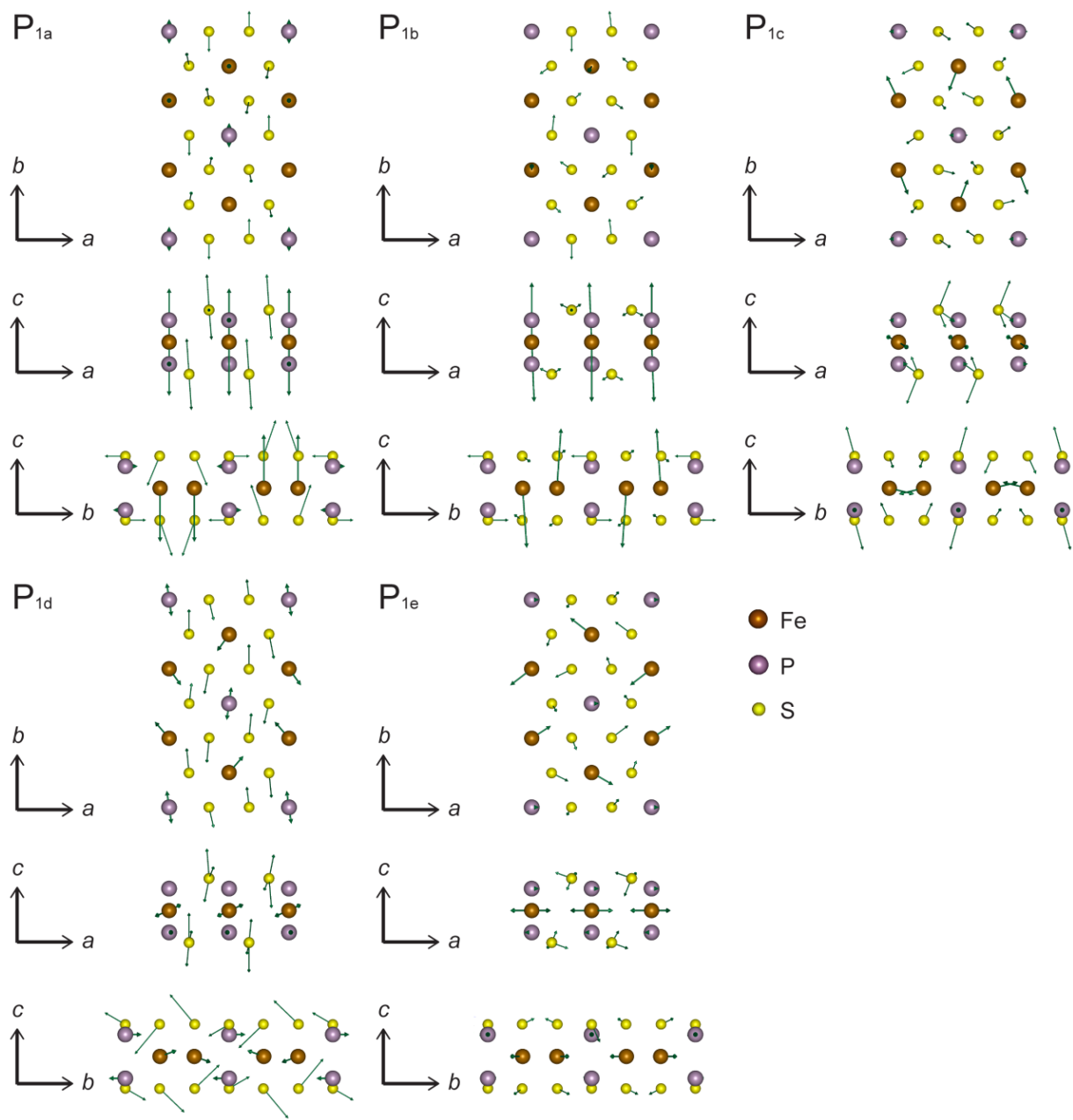

**Figure S1.** Calculated lowest-frequency Raman active modes.

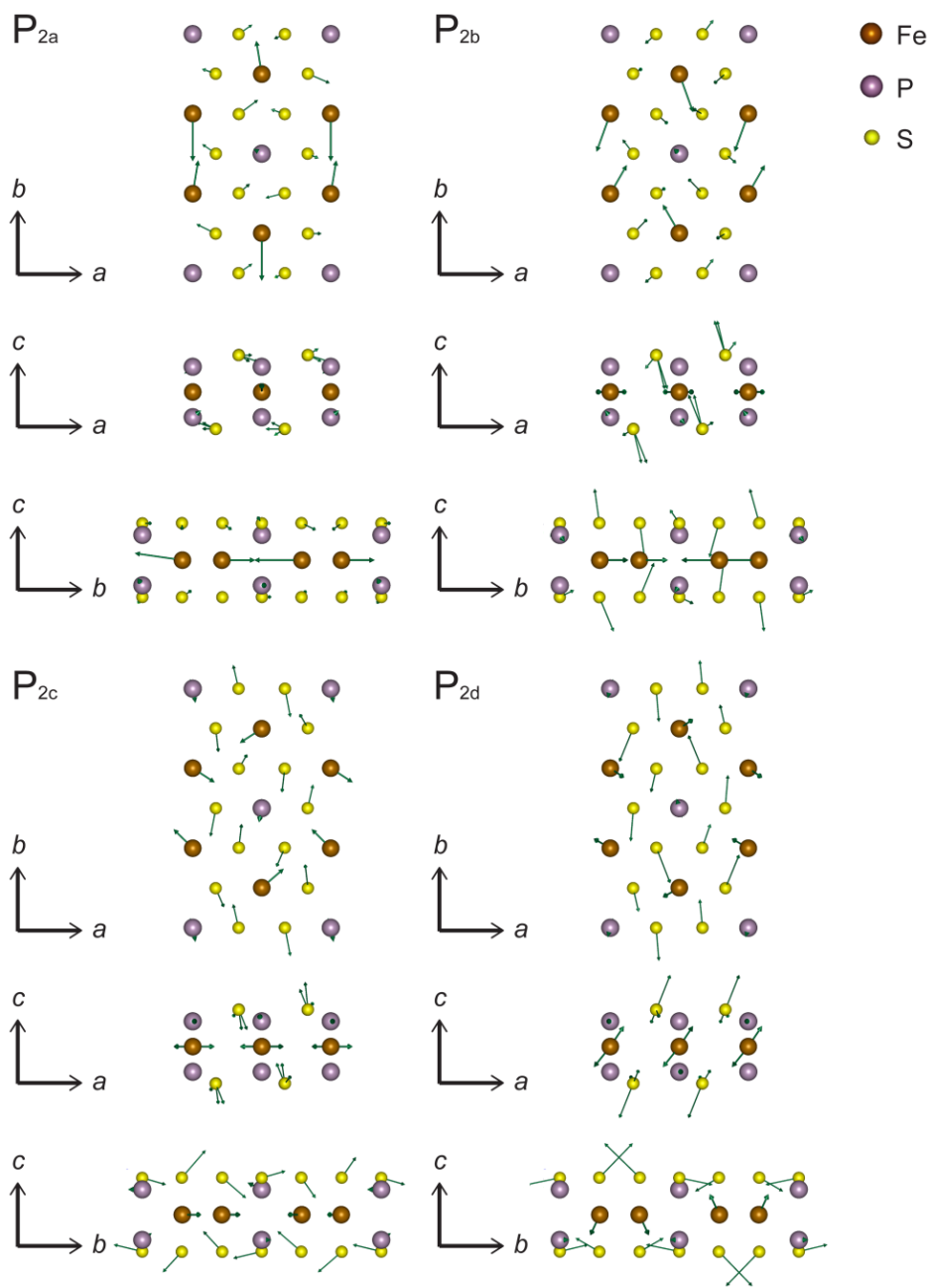

**Figure S2.** Calculated Raman active modes corresponding to $P_2$.

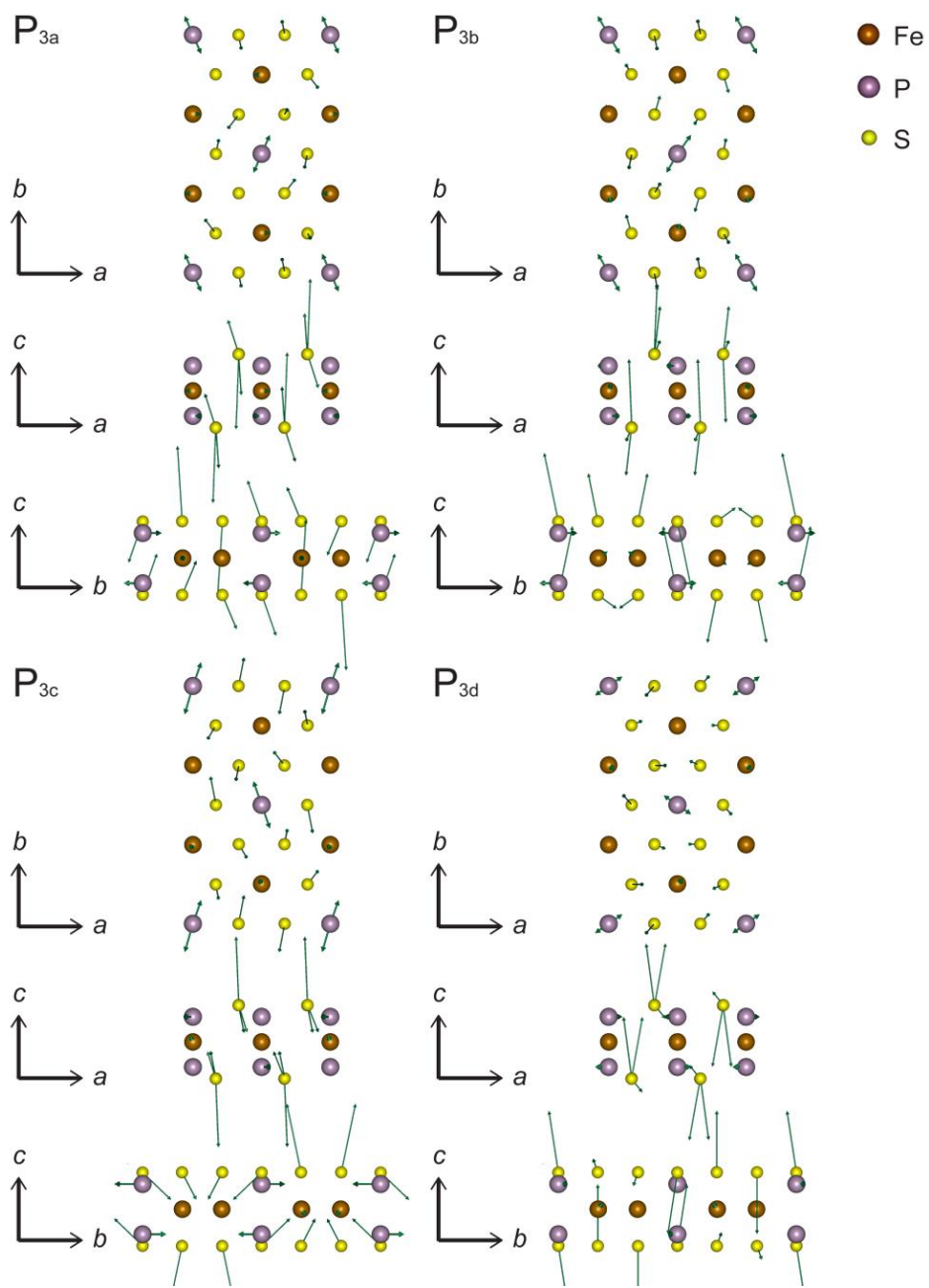

**Figure S3.** Calculated Raman active modes corresponding to $P_3$.

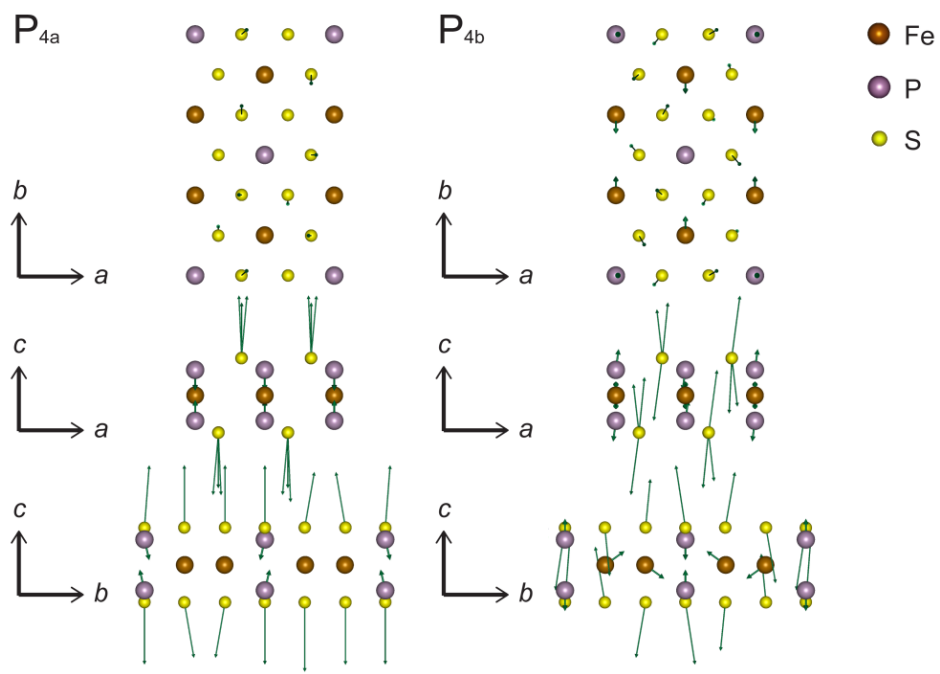

**Figure S4.** Calculated Raman active modes corresponding to $P_4$.

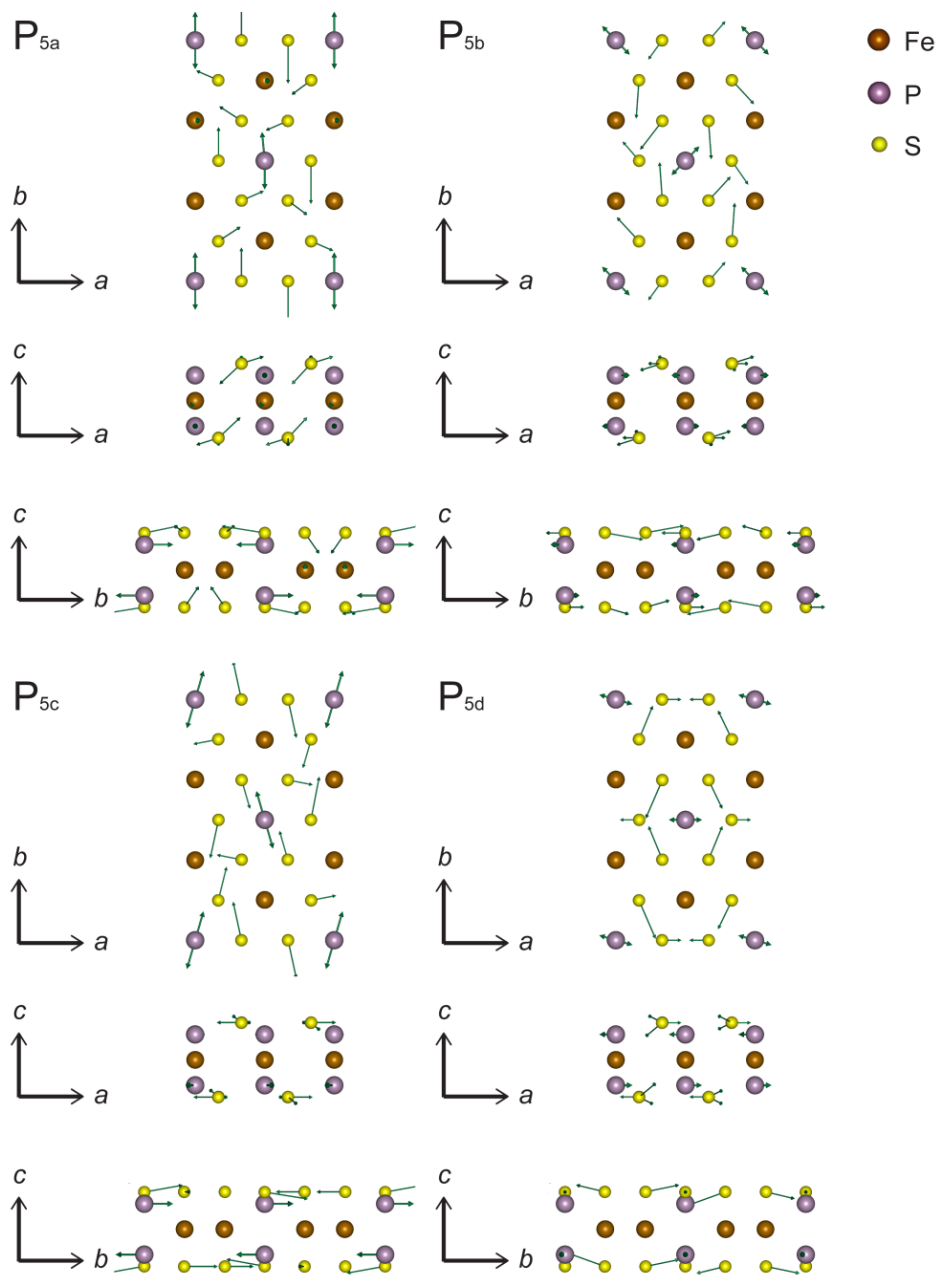

**Figure S5.** Calculated Raman active modes corresponding to P$_5$.

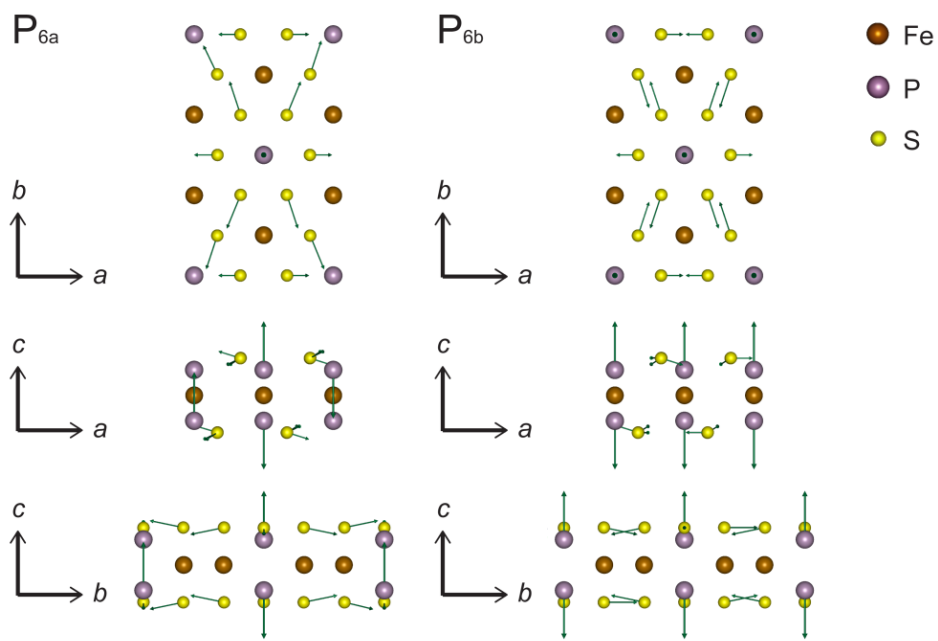

**Figure S6.** Calculated Raman active modes corresponding to $P_6$.

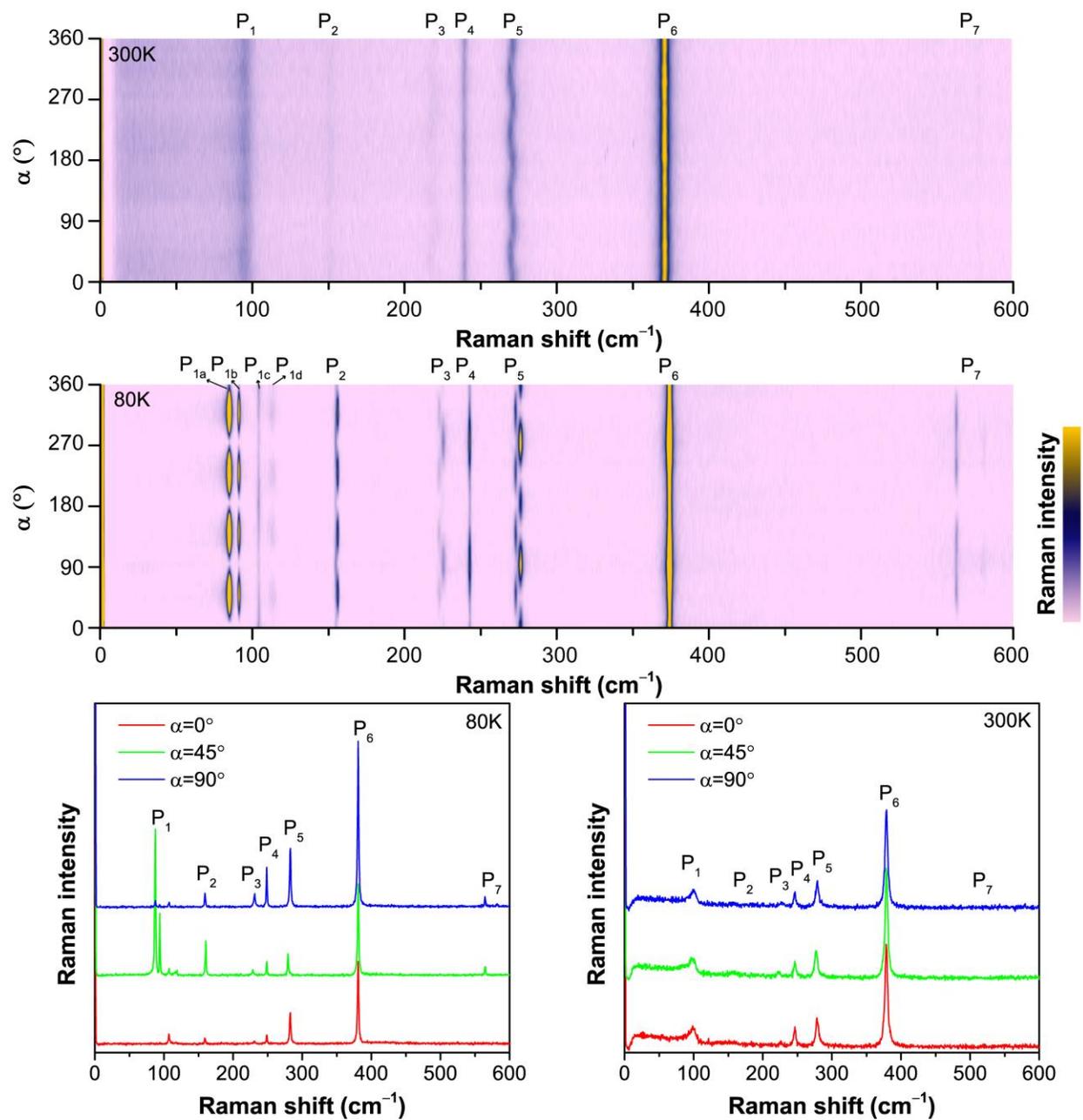

**Figure S7.** Polarized Raman spectra of bulk FePS$_3$ in the extended spectral range.

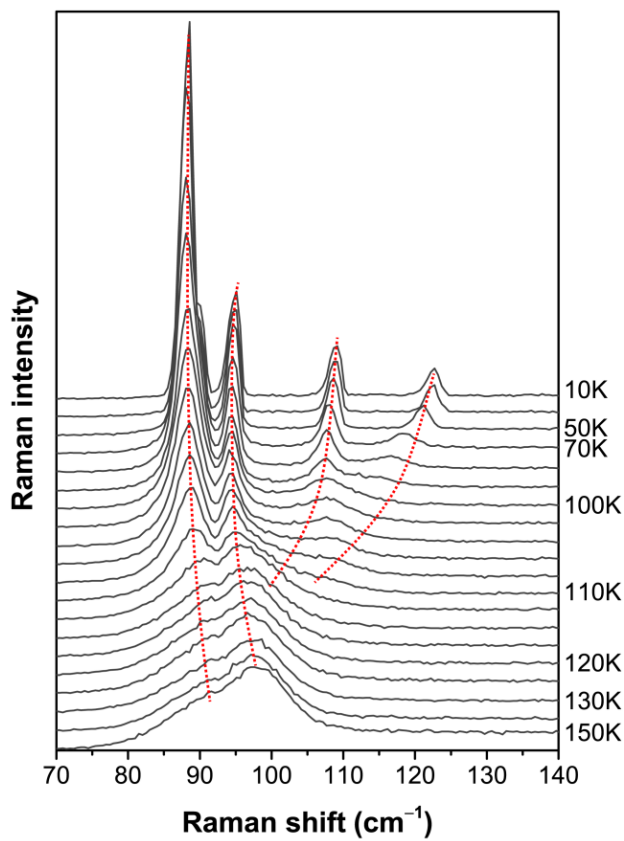

**Figure S8.** Temperature dependence of $P_1$ components in bulk $FePS_3$. Red dashed curves are guide to eyes.

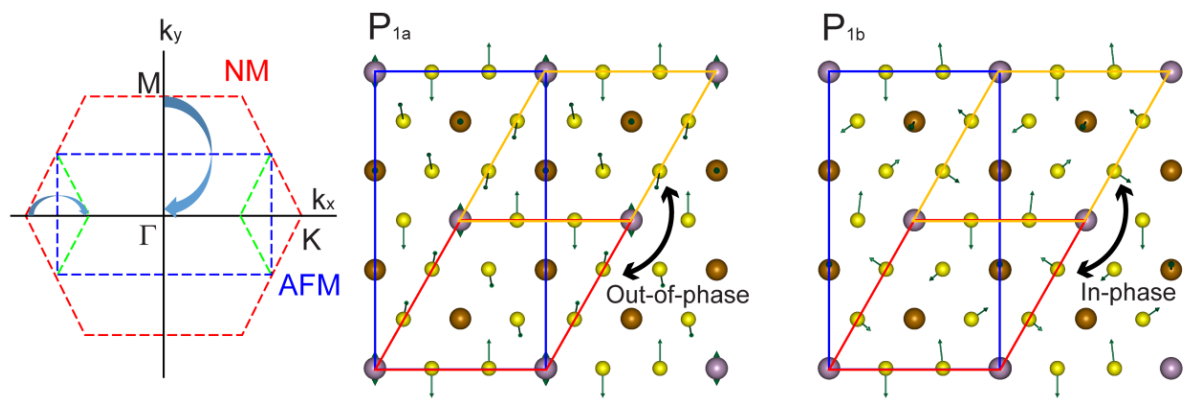

**Figure S9.** Schematics and examples of zone-folding effect. Blue lines correspond to the Brillouin zone and the unit cell in the antiferromagnetic phase (AFM), and red lines for the non-magnetic phase (NM).

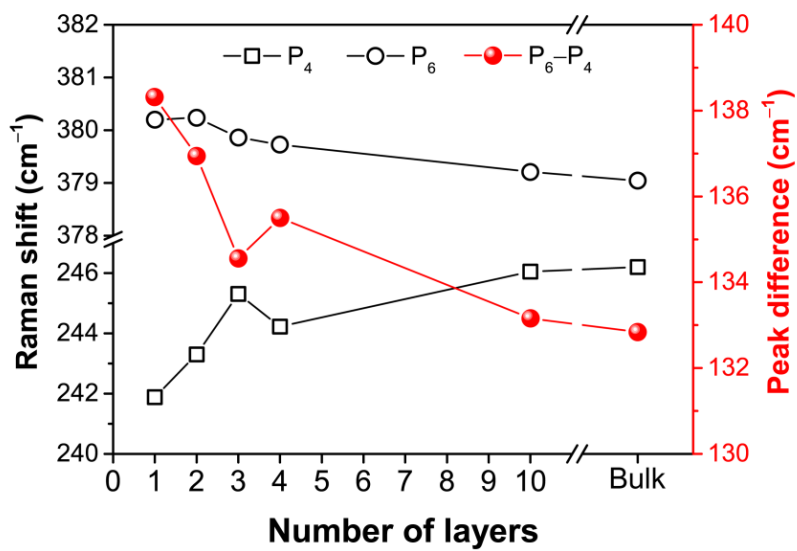

**Figure S10.** Peak positions of P$_4$ and P$_6$ and their difference (P$_6$−P$_4$) as a function of thickness.

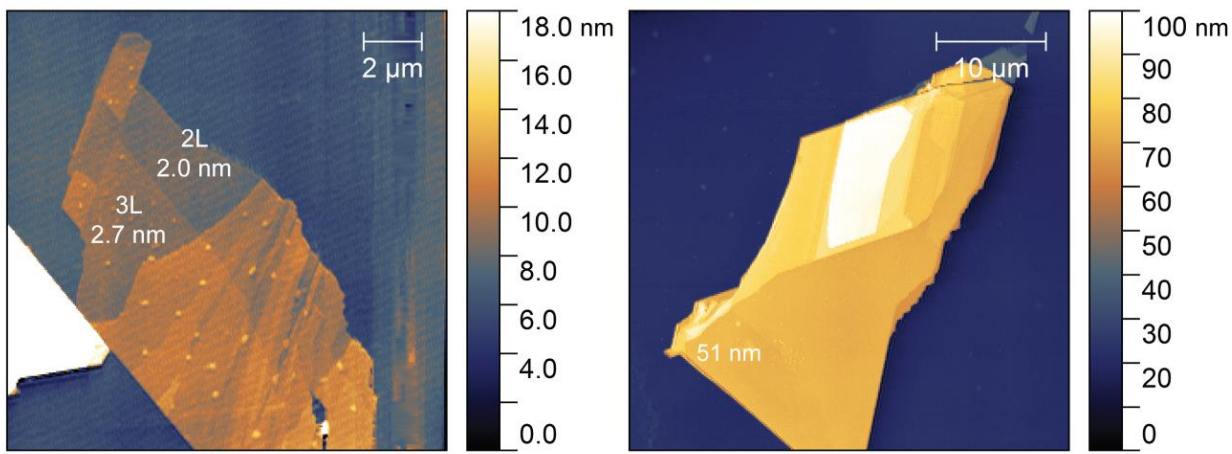

**Figure S11.** Representative atomic force microscopy images of exfoliated FePS$_3$.

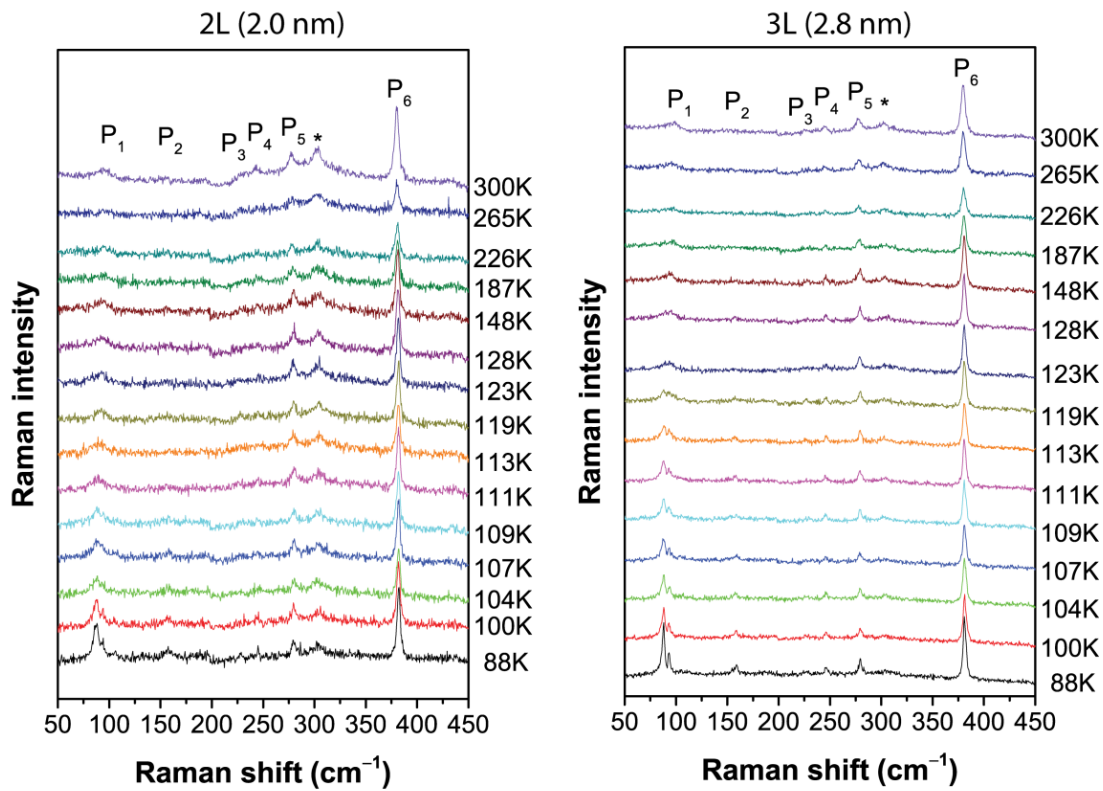

**Figure S12.** Temperature dependence of Raman spectra for 2L and 3L samples. The peak at ~300 cm$^{-1}$ marked by * is a signal from the Si substrate.

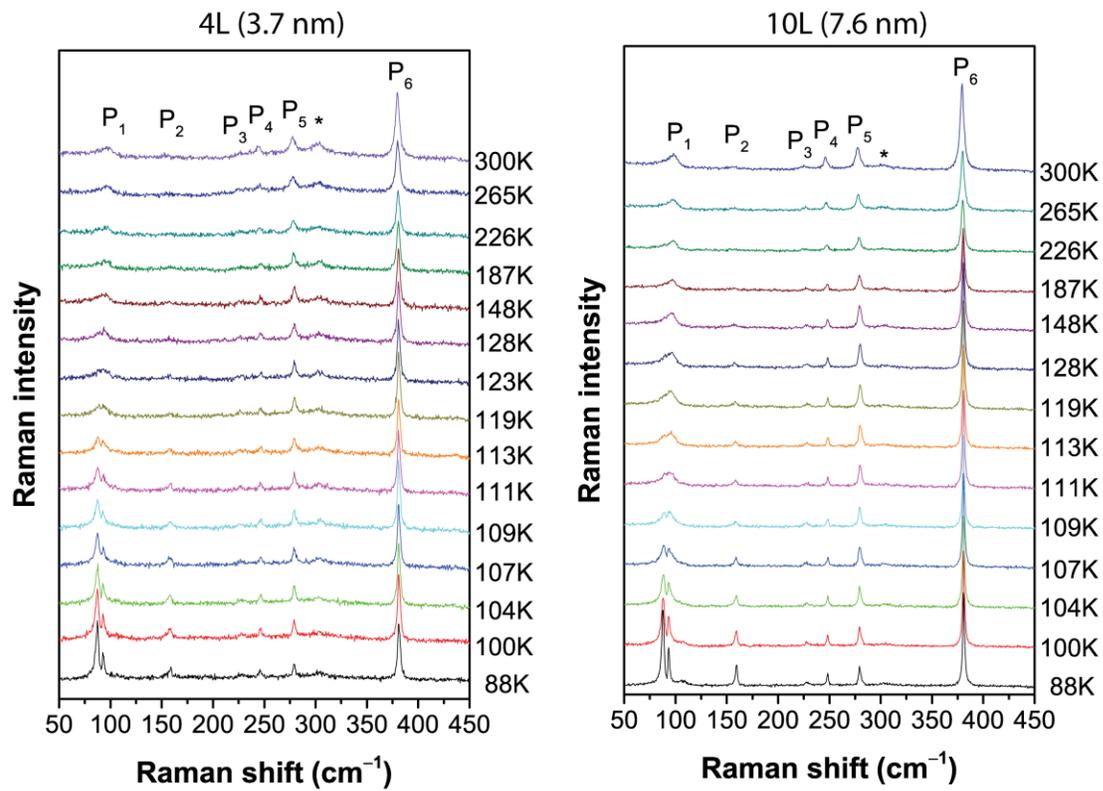

**Figure S13.** Temperature dependence of Raman spectra for 4L and 10L samples. The peak at ~300 cm$^{-1}$ marked by * is a signal from the Si substrate.

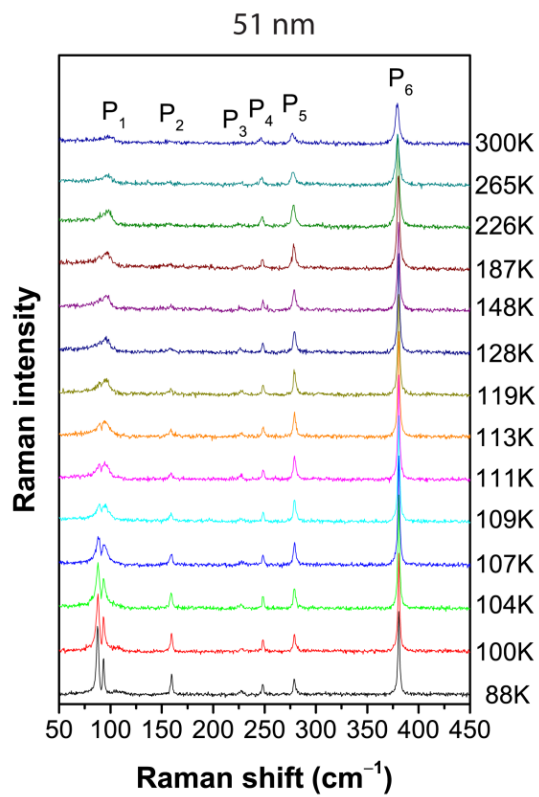

**Figure S14.** Temperature dependence of Raman spectrum for 51-nm-thick sample.

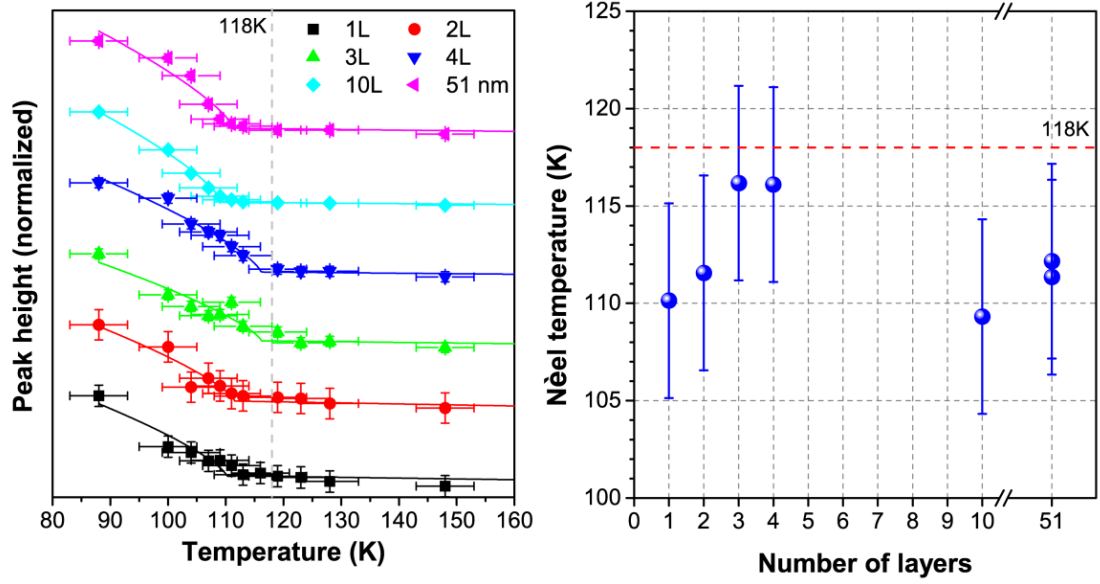

**Figure S15.** Temperature dependence of $P_{1a}$ peak height with different thicknesses. The fitting curves are shown as solid lines. The curves are vertically stacked for clarity. Obtained Néel temperature as a function of thickness.

**Fitting model for temperature dependent Raman intensities**

Suzuki et al.[1] developed a general theory of the spin-dependent phonon Raman scattering in magnetic crystals as follows.

$$S(T) = (n_0 + 1)\left| R + M \langle S_0 \cdot S_1 \rangle / S^2 \right|^2,$$

where $n_0$ is the Bose-Einstein factor, $R$ and $M$ are coefficients of the spin independent and dependent parts, respectively, and $\langle S_0 \cdot S_1 \rangle / S^2$ is the nearest neighbor spin correlation function. In their categorization, FePS$_3$ belongs to the $0 < R/M$ case. Near the phase transition temperature, the spin correlation function can be approximated as $\langle S_0 \cdot S_1 \rangle / S^2 \sim m^2(T) \sim (T_c - T)^{2\alpha}$.[2,3] By assuming that the critical exponent is the same regardless of thickness, we obtained $\alpha \sim 0.37$ which is close to the typical value. In Figure S15, we obtained the Néel temperature by fitting this function to the data at low temperature.

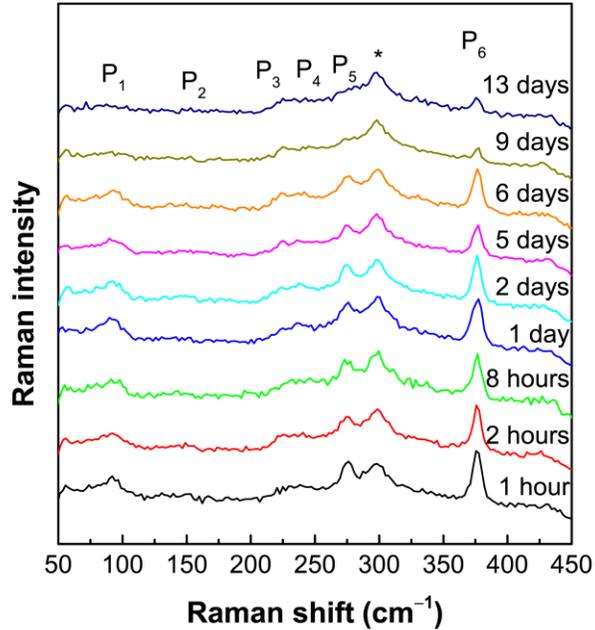

**Figure S16.** Degradation of 2L FePS$_3$ exposed in the ambient condition estimated from Raman spectrum. The sample is stable up to 6 days. The peak at ~300 cm$^{-1}$ marked by * is a signal from the Si substrate.

**II. Effect of symmetry on calculated atomic structure and phonons of FePS$_3$**

It is widely accepted that bulk FePS$_3$ in the non-magnetic phase has a mirror symmetry (mirror plane containing the *a* and *c* axes in Fig. 1 of the main manuscript).[4] However, according to our DFT+*U* calculations on monolayer FePS$_3$ assuming anti-ferromagnetic ground state, the atomic structure without this mirror symmetry is always lower in energy than the one with it (we have checked both the cases $U = 4.2$ eV and $U=6.0$ eV). Similar breaking of the honeycomb lattice symmetry of FePS$_3$ monolayer was reported from a first-principles DFT+$U$ study.[5] The results presented in Figures S1–S6 are for this case without mirror symmetry. Interestingly, similar calculations assuming a non-magnetic ground state preserves mirror symmetry. (We discuss more on this issue later when we discuss the results of our calculations as well as other calculations in the literature on bulk FePS$_3$.) Although this breaking of the mirror symmetry is rather small (Fe atoms being displaced by 0.04 angstroms), it affects some phonon modes appreciably. Since the point group (the origin being at the center of the two P atoms at the same position in the *ab* plane) is reduced from $C_{2h}$ to $C_i$, each phonon mode is classified according to the irreducible representations of $C_i$: inversion-even modes (Raman-active) and inversion-odd modes (Raman-inactive).

On the other hand, one cannot exclude the possibility that the actual antiferromagnetic phase has a structure with mirror symmetry ($C_{2h}$) as in the non-magnetic phase. In order to see if this possibility would impact our interpretation, we carried out additional calculations in which we constrained the point group to $C_{2h}$. Figure S17 shows the Raman-active phonon modes of monolayer FePS$_3$ with the lowest frequencies, calculated in this way. Although there is an unstable phonon mode corresponding to the breaking of the mirror symmetry (Figure S18), we note that the first and the second modes in Figure S17 are nearly identical to the first and the second Raman-active modes of antiferromagnetic monolayer FePS$_3$ ($C_i$) shown in Figure S1. Therefore, regardless of whether this reduction of the symmetry occurs in monolayer or not, the conclusion that our calculation explains the zone-folding behavior in Raman spectrum is solid.

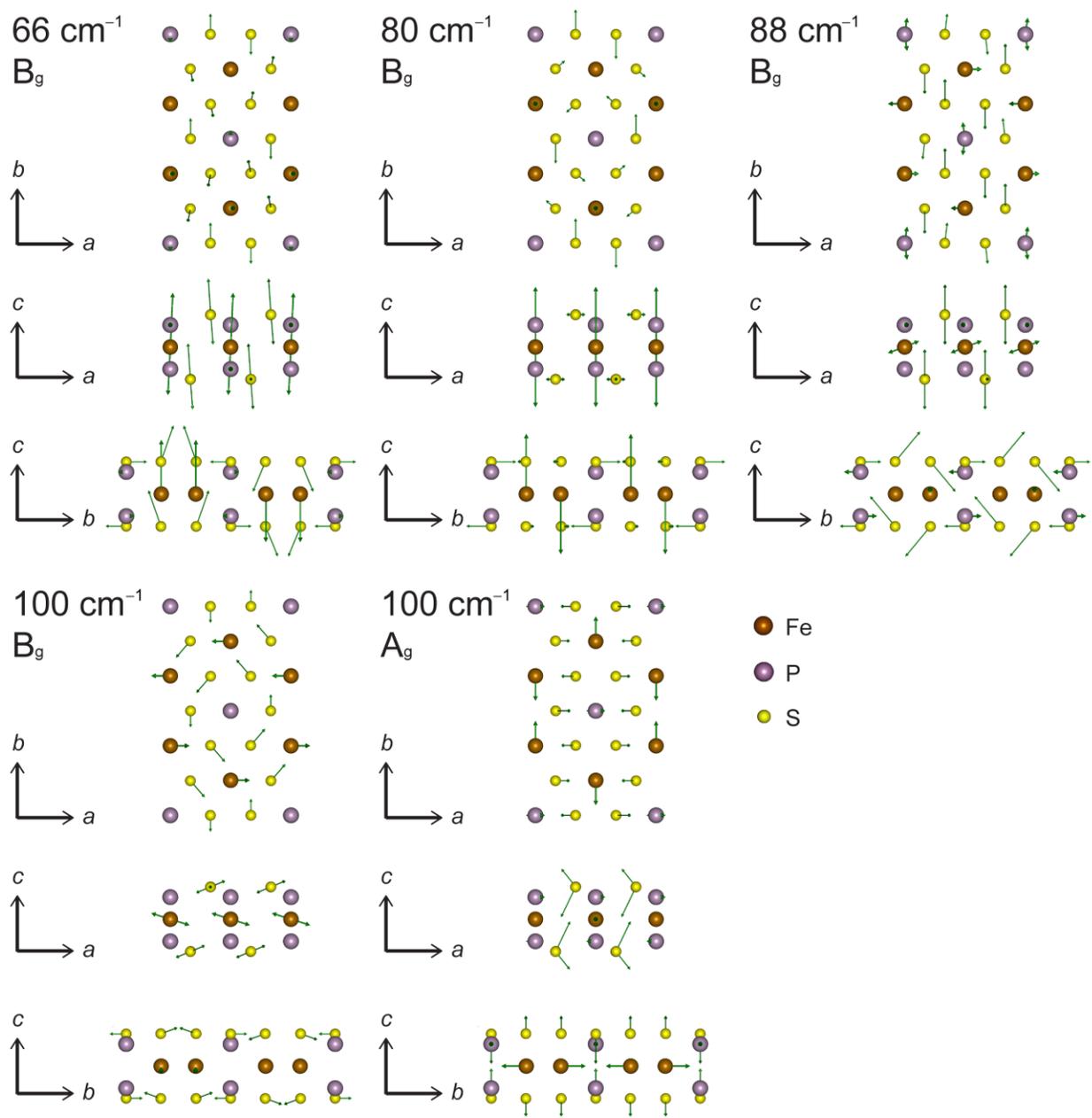

**Figure S17.** Lowest-energy stable phonon modes of monolayer FePS$_3$; the point group is constrained to $C_{2h}$.

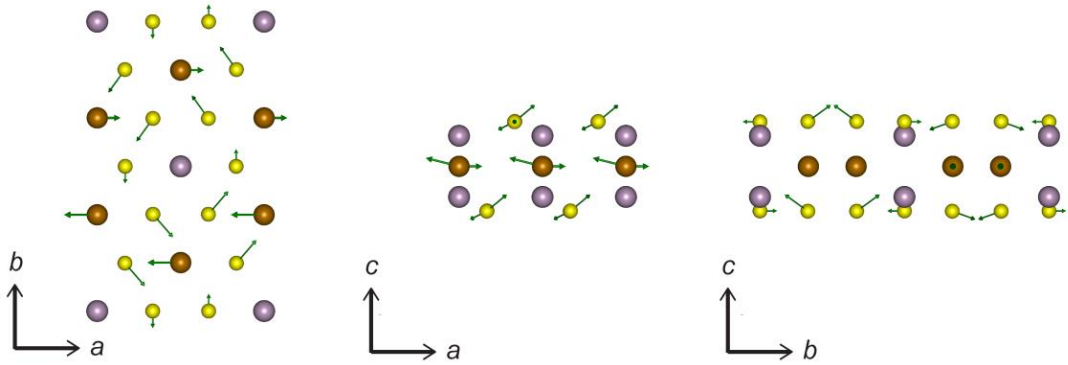

**Figure S18.** Unstable phonon mode of monolayer FePS$_3$ when the point group is constrained to $C_{2h}$.

**Calculated atomic structure and phonons of bulk FePS$_3$**

Although we mainly presented the results of calculation on monolayer FePS$_3$ in the main manuscript, the corresponding results on bulk crystal are similar. First, we note that a slight breaking of the $C_{2h}$ mirror symmetry leads to a lower-energy structure if antiferromagnetic ground state is assumed and first-principles DFT+$U$ method is employed. The two lowest phonon modes ($P_{1a}$ and $P_{1b}$) in monolayer FePS$_3$ still manifest themselves in the phonon spectrum of bulk FePS$_3$. Since the unit cell is doubled along the $c$ direction, each monolayer mode is also doubled. These results show that the analysis of the phonon modes in monolayer FePS$_3$ is sufficient for discussing the zone-folding behavior in Raman spectrum.